\documentclass[twocolumn]{revtex4}

\usepackage{graphicx}
\usepackage{graphics}
\usepackage{amssymb}
\usepackage{hyperref}
\usepackage{latexsym}

\def\be{\begin{equation}}
\def\ee{\end{equation}}
\def\bea{\begin{eqnarray}}
\def\eea{\end{eqnarray}}

\begin{document}
\title{Polymorphism, thermodynamic anomalies and network formation in an atomistic model with two internal states}
\author{Chandana Mondal and Surajit Sengupta}
\address{Centre for Advanced Materials, Indian Association for the Cultivation of Science, Jadavpur, Kolkata-700032, India}

\begin{abstract}
Using molecular dynamics simulations we study the temperature-density phase diagram of a simple model system of particles in two dimensions. In addition to translational degrees of freedom, each particle has two internal states and interacts with a modified Lennard-Jones potential which depends on relative positions as well as the internal states. We find that, despite its simplicity, the model has a rich phase diagram showing many features of common network-forming liquids such as water and silica, including polymorphism and thermodynamic anomalies. We believe our model may be useful for studies concerning generic features of such complex liquids. 
\end{abstract}

\maketitle

\section{Introduction}
\noindent
Water\cite{water1}, molten silicon\cite{silicon} and silica glasses\cite{silica} are common examples of network-forming liquids which are distinguished from simple fluids (like liquid argon) by the presence of complex intermolecular interactions allowing bond-formation only in favoured directions. Similar ability is also seen in macro-molecular systems
 such as colloidal particles with patchy interactions\cite{Sciortino1,Sciortino2}. Due to their propensity
 toward forming extensive networked structures, especially in supercooled states, such liquids show many
 intriguing properties such as isobaric density maxima, negative thermal expansion coefficients($\alpha_{p}$),
 and increase in the isothermal compressibility($\chi_{T}$) and isobaric heat capacity($c_{p}$) upon
 supercooling. Also, network-formers exhibit polymorphism i.e. existence of
 multiple crystalline as well as amorphous states varying in local coordination and density\cite{polymorph}.
 Over the years a lot of theoretical, computational and experimental effort has been made in order to
 understand the origin of such anomalous properties of network-formers especially
 water\cite{anomaly1, anomaly2, anomaly3, anomaly4, anomaly5, anomalous}. It is
 now well established from a microscopic point of view that the anomalous properties of water are related to
 the existence of high and low density forms\cite{water1, water2, water3} of the supercooled liquid
 in addition to the several stable states of ice. The metastable liquid-liquid critical
 point\cite{liquid-liquid,-vepressure1} in water, buried deep within the (stable) ice phase can, nevertheless,
 crucially influence the dynamics of nucleation and amorphization. A very similar scenario has recently been
 proposed for liquid silicon\cite{-vepressure2} which suggests that thermodynamical anomalies may be expected
 for any substance which has low, as well as high, density crystalline states. Since low energy open
 crystals are most easily (though not exclusively) formed in systems of molecules with highly directional
 bonding e.g. the hydrogen bond network in water, strongly orientation dependent interactions appear to
 be a prerequisite for water-like properties\cite{lattice}.  

While the connection between directional bonding, polymorphism, liquid-liquid phase transitions and thermodynamic anomalies is established, the univeral nature of this connection opens up many new avenues of investigation. What is the effect of confining fields and substrates on the properties of networked liquids? What happens to the network driven under shear flow? How does one characterize the dynamics of network formation? To answer these, and  other similar questions one ideally needs a simple and generic model which is amenable to simulation on the computer, as well as accessible to fairly simple theoretical analyses. Though such generic models do exist, they are either idealized, as in a lattice model\cite{lattice, lattice-model}, or not much simpler to handle than fully realistic water (or silica) models\cite{Mercedes-Benz}; the most time consuming part being an accurate treatment of molecular rotations and translation-rotation coupling at high densities\cite{ums}.

In this paper, we describe a simple and generic {\em atomistic} model system which shows many of the characteristics of network forming liquids without the complication of molecular rotations. Our model consists of particles $i = 1 \dots N$ which have an internal coordinate $S_i$.  Unlike an angular coordinate, $S_i$ are discrete and can take only one of two possible values $\pm 1$. The interaction between particle pairs $i,j$ is strongly directional and depends on the displacement ${\bf r}_{ij}$ as well as the values of $S_i$ and $S_j$. The internal state $S_i$ may be thought of as mimicking rotations though it is not necessary to do so.

\begin{figure}[ht]
\begin{center}
\includegraphics[width=8.0cm]{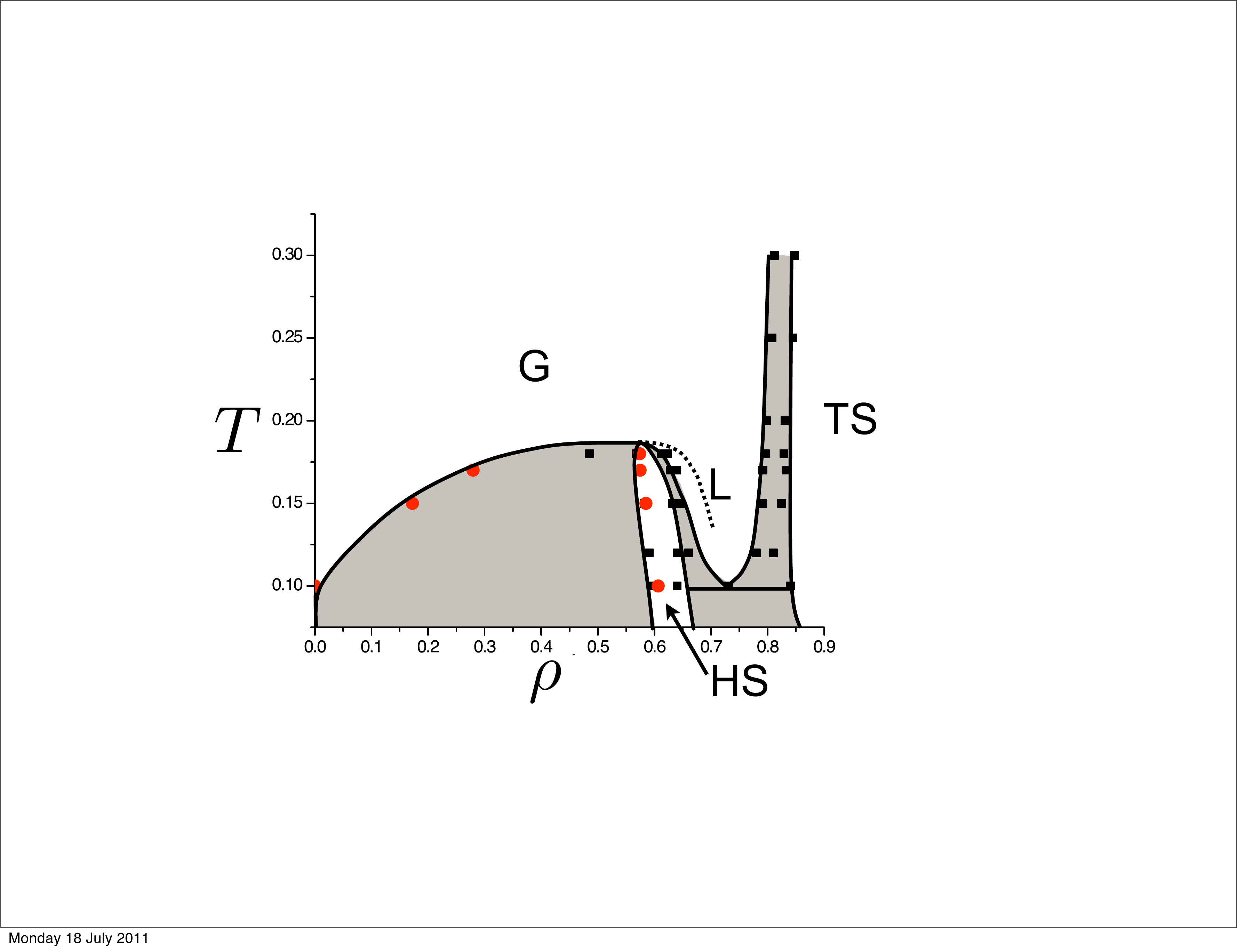}
\end{center}
\caption{(color-online)~Complete phase diagram of the system in the $\rho-T$ plane as obtained from our simulations. The red circles, obtained from block-analysis technique, show the gas (G) and hexagonal-solid (HS) densities at different temperatures. The black squares, obtained from pressure-density isotherms, show the phase boundaries of HS-liquid (L), HS-triangular-solid(TS) and L-TS coexistance regions. The solid lines joining the red circles and black squares are guides to the eye. The two phase coexistence regions are shaded in gray. Also shown is the locus of the pressure-minima as a dotted line.}
\label{Fig.1}
\end{figure}

\begin{figure}[ht]
\begin{center}
\includegraphics[width=6cm]{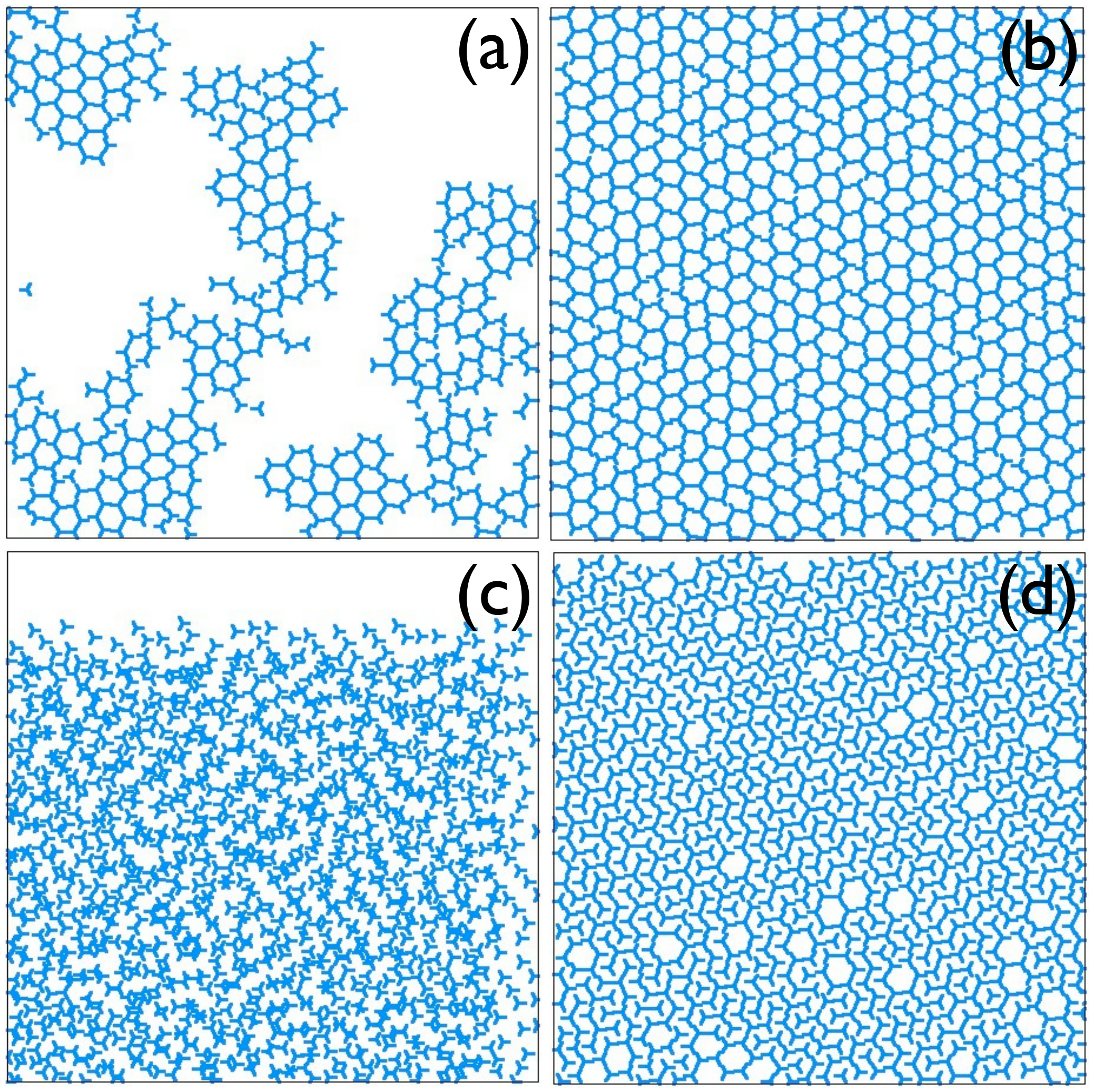}
\end{center}
\caption{~Snapshot picture after $10^{7}$ MD-steps for (a) $\rho = 0.3,
 T = 0.1$  showing the G-HS co-existence, (b) $\rho = 0.6, T = 0.1$ showing the
 HS phase, (c) $\rho = 0.7, T = 0.3$ showing the L phase, (d) $\rho = 0.86,
 T = 0.1$ showing  HS-TS co-existance.}
\label{Fig.2}
\end{figure}

Our main results are the following. Despite the simplicity of the model, we obtain a rich phase diagram (Fig.\ref{Fig.1}) using finite-size scaled block analysis technique\cite{Block}. The system has two distinct crystalline phases viz. a low density honeycomb lattice (HS) and a closed packed triangular crystal (TS). These crystals melt into a liquid (L) or sublimate to a vapor (G) under appropriate conditions (Fig.\ref{Fig.2}(a)-(d)). Remarkably, the HS phase coexists with a liquid which is of {\em higher} density than the solid; very similar to water and (normal) ice at atmospheric temperatures and pressures. The liquid phase at low temperatures shows anomalous variation of pressure and density with temperature similar to that of water and silicon. Correlations in the liquid phase\cite{hansen} begin to show strong directional modulations before such anomalies arise pointing to the formation of a dynamic network within the liquid. Because of the simplicity of our model, such correlations can easily be obtained from perturbative, liquid state integral equation theory.


This paper is organized as follows. In Section 2 we introduce the essential features of the model and provide
 details of our simulations. We discuss our results focusing on the complete density-temperature
 phase-diagram of the system, anomalous properties and the development of short-range order in Section 3.
We conclude in Section 4 pointing out some directions of future work.

\section{The model and simulation details}

\subsection{Interaction potential}

Our system consists of $N$ particles in a box of area $A$, with $50:50$ mixture of particles with the internal coordinates $S_i = \pm 1$  interacting with each other through a simple pairwise-additive 2-body interaction in two dimensions. The interaction potential between particles $i$ and $j$, separated by the radius vector ${\bf r}_{ij}$, viz. $U_{S_iS_j}({\bf r}_{ij})$ may be decomposed as follows:
\begin{eqnarray}
U_{1,1}(r) = U_{-1,-1}(r) & = & V_{0} (r) \nonumber \\
U_{1,-1}(r,\theta) & = & V_{0}'(r) + \delta V(r,\theta) \nonumber \\
U_{-1,1}(r,\theta) & = & V_{0}'(r) - \delta V(r,\theta)
\end{eqnarray}
where $r = |{\bf r}_{ij}|$ is the magnitude and $\theta = \cos^{-1} ({\bf r}_{ij}{\bf \cdot}\hat{i}/r_{ij}) $ the angle ${\bf r}_{ij}$ makes with the $x$-axis. The isotropic ($\theta$ independent) parts of the potential are given by,
\begin{eqnarray}
V_{0}(r) & = & U_{LJ}(r)+\epsilon, \,\,\ r \leq r_{min.} \nonumber \\
 & = & 0, \,\,\,\,\,\,\,\,\,\,\,\,\,\,\,\,\,\ r > r_{min.}  
 \end{eqnarray}
with $U_{LJ}(r)  = 4 \epsilon [(\sigma/r)^{12} - (\sigma/r)^6]$, the usual $6-12$ Lennard-Jones interaction; $r_{min.} =2^{\frac{1}{6}}\sigma$ and $\epsilon$ is the position and depth of the minimum in $U_{LJ}$ respectively and,
\begin{eqnarray}
V_{0}'(r) & = & V_{0}(r) + V_{1}(r) \nonumber \\
\end{eqnarray}
where,
\begin{eqnarray}
2V_{1}(r) & = & -\epsilon-E_{cut}, \,\,\,\,\ r \leq r_{min.} \nonumber \\
& = & U_{LJ}(r)-E_{cut}, \,\ r>r_{min.} 
\end{eqnarray}
with $E_{cut}=U_{LJ}(r_{cut})$ with $r_{cut}^{2}=2.5\sigma^{2}$. Finally, the angle dependent part,
\begin{eqnarray}
\delta V(r,\theta) & = & V_{1}(r) \cos(3 \theta)
\end{eqnarray}

\begin{figure}[ht]
\begin{center}
\includegraphics[width=8cm]{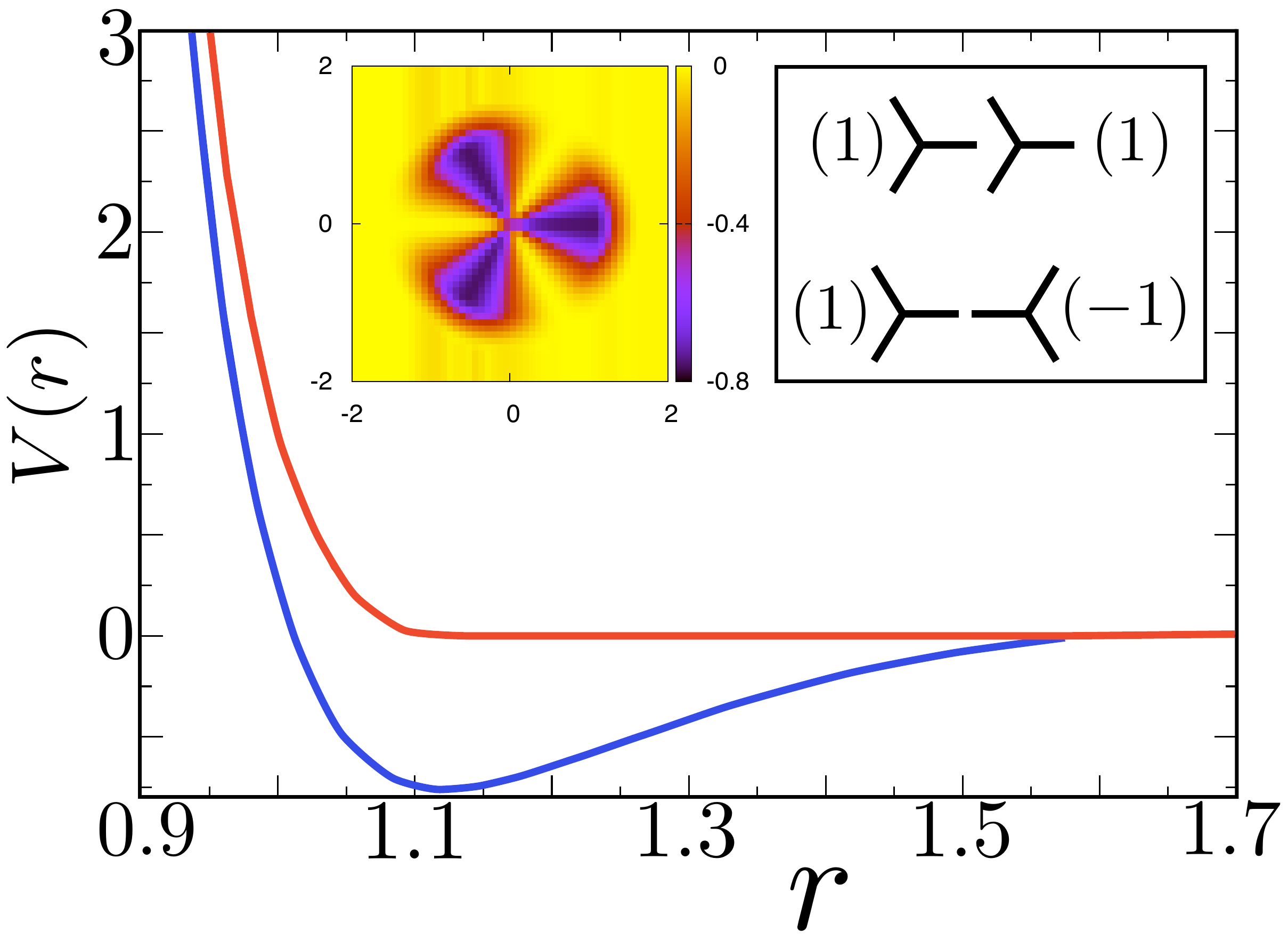}
\end{center}
\caption{(color-online)~Plot of the interaction potential between two particles with same: red/light grey curve $(1,1)$, and opposite: blue/dark grey curve $(1,-1)$ values for $(S_i,S_j)$. The plots are along the  $+ve$ x-axis where the potential has the deepest minimum. Inset top left: Polar plot of the angular part of the potential for $S_i = 1$ and $S_j = -1$ showing the three directions along which the potential has minima. Inset top right: the geometries for the two distinct cases are pictorially illustrated.}
\label{Fig.3}
\end{figure}

In Fig.\ref{Fig.3} we have plotted the full interaction potential as a function of $r_{ij}$ for two cases, namely when $S_i = S_j = 1$ (red/light grey curve) and $S_i = -S_j = 1$ (blue/dark grey curve) along the direction $\theta_{ij} = 0$. The form of the potential ensures that an interchange of the labels $i$ and $j$ is accompanied by a rotation of the potential by $\pi$.
The units for distances and energies are the same as in a Lennard-Jones system, viz. $\sigma$ and $\epsilon$ and for simplicity we have taken $\sigma = \epsilon = 1$ without loss of generality.

\subsection{Details of the simulations}

We perform molecular dynamics simulation\cite{ums} with $N = 1176$ particles in a two-dimensional rectangular box of area A using a standard velocity~-verlet algorithm taking care to choose an integration time step  $\Delta t = 10^{-4}$ such that the total energy(E) is conserved to $1$ in $10^6$ or better\cite{torque}. We perform equilibration runs in constant N,A,T ensemble, while production
 runs are performed in the constant N,A,E ensemble.The temperature is fixed by rescaling the velocities and
 thermodynamic quantities are obtained after equilibration starting from either a honeycomb or a triangular
 crystal depending on the density. Periodic boundary conditions are applied in both directions. In each state,
 we equilibrated the system for $6\times10^{6}$ time steps and then calculated time-averages for
 $4\times10^{6}$ time steps. 
 
 We obtain most of the phase diagram shown in Fig.\ref{Fig.1} using the block analysis technique\cite{Block}
 which has been used in the past for similar systems and has been described in detail. Briefly, the simulation
 box is divided into $M_b$ equal sized blocks and the local density is evaluated for each block. From the block
 densities one can construct the block density probability distribution $P_{M_b}(\rho)$ for every block
 indexed by $M_b$. In a single phase region, this distribution function is a simple Gaussian centered about the
 mean density $\rho$ with a width which depends on $M_b$ and the isothermal compressibility of the system
 $\chi_T$. In a co-existence region, the distribution function consists of a sum of two Gaussians each centered
 on the densities of the co-existing phases, which are then obtained by fitting a sum of appropriately chosen
 Gaussians and extrapolating the resulting co-existence densities to the thermodynamic limit $M_b \to 0$. In
 order that the block-analysis technique works, the co-existing densities need to be well separated. The block
 analysis technique therefore works best for liquid~-gas and solid~-gas coexistence. For parts of the phase
 diagram where the co-existence regions are narrow, we obtain co-existing densities in the usual way from
 pressure isotherms using a Maxwell's equal-area construction. Apart from co-existence densities, we 
 also calculate pair distribution functions for like and unlike $S_i$ to characterize the various phases.

\begin{figure}[ht]
\begin{center}
\includegraphics[width=7.0cm]{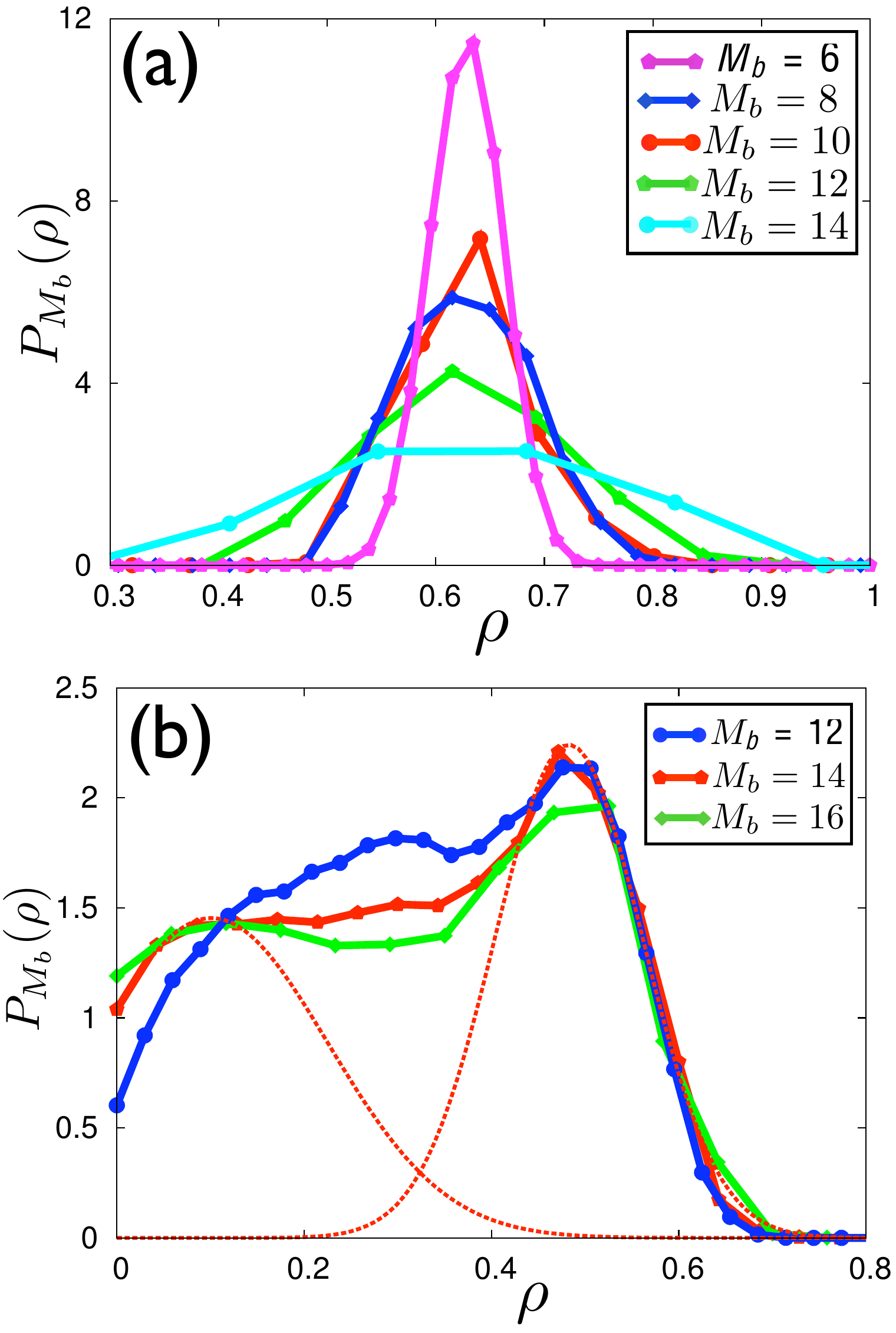}
\end{center}
\caption{(color-online)~The probability distribution curves for density $P_{M_b}(\rho)$ for a few values of $M_b$ in the (a) in the HS region($\rho = 0.628$, $T = 0.12$) and (b) G-HS co-existence region ($\rho = 0.35$, $T = 0.15$). The lines joining the points are guides to the eye. In (b), we have included two Gaussians (dashed red curves) which were fitted to the data for $M_b = 14$ }
\label{Fig.4}
\end{figure}

\section{Results and discussions}

\subsection{Phase-Diagram}
We obtain the density-temperature phase diagram for this system using the block-analysis\cite{Block} technique along with finite-size scaling described above. In Fig.\ref{Fig.4} (a) we show the density distribution curves for various values of $M_b$ in a region where a single homogeneous phase (the honeycomb solid, HS) is stable. Each of the curves may be represented by single Gaussians with the curves becoming sharper as $M_b$ decreases as expected. Since we work in the constant density ensemble, the corresponding distribution function for $M_b = 1$ is a trivial delta function at $\rho$. In contrast, when the system is in a two~-phase region, we expect the density distribution to be bi-modal whenever the size of the blocks are comparable to the size of the heterogeneities. This is illustrated in Fig.\ref{Fig.4} (b) where we have obtained $P_{M_b}(\rho)$ for G-HS coexistence. For the largest block size shown i.e. for $M_b=12$, the distribution is still uni-modal, although it develops a prominent shoulder. For $M_b \gtrsim 14$, however, the two peaks can be clearly resolved which gives us the coexisting gas and solid densities.

\begin{figure}[ht]
\begin{center}
\includegraphics[width=6.0cm]{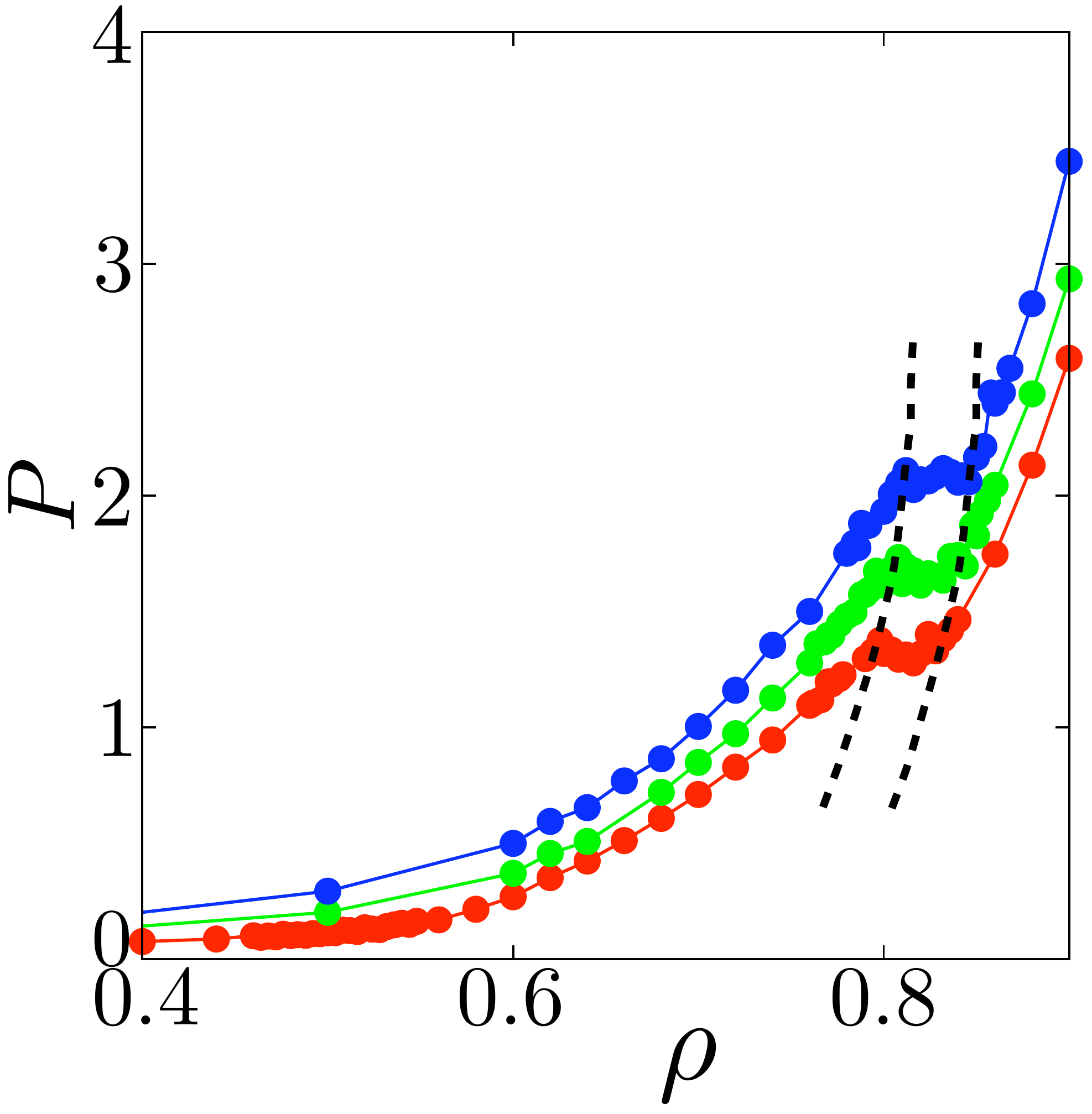}
\end{center}
\caption{(color-online)~$P(\rho)$ isotherms along $T = 0.2$ (red), $.25$  (green) and $.3$(blue). The flat region is the L-TS coexistence region. The estimated coexistence densities lie on the dashed curves as shown.}
\label{Fig.5}
\end{figure}

It is possible that the difference in the coexisting densities is so small that the bi-modal structure of the density distribution cannot be resolved. This happens when the coexisting phases are liquid and solid or two different solids. In this case, as mentioned, we use pressure isotherms to determine the coexistence densities. The pressure isotherms across the L-TS coexistence region are shown in Fig.\ref{Fig.5} for three temperatures. The computed coexistence densities lie along the dashed curves as shown.

The complete density-temperature phase-diagram is shown in Fig.\ref{Fig.1} and snapshot pictures of a few of the featured phases are shown in Fig.\ref{Fig.2}. Despite the relative simplicity of the interactions, our system shows a rather complex phase behavior. Firstly, there are two very different crystalline structures, the honeycomb and triangular solids, HS and TS. While the former has an open structure with strong directional bonding, the latter is close packed. The HS solid can coexist with either a low density gas, G, or with a liquid, L, which has a density higher than itself. This unique feature is reminiscent of many networked liquids such as water. The triangular solid TS, can be in coexistence with L or even with another solid viz. HS. Again this is a feature common in the water-ice system where various low and high density forms of ice may coexist with each other for appropriate values of the pressure.

\subsection{Thermodynamic  anomalies}
\begin{figure}[ht]
\begin{center}
\includegraphics[width=9.0cm]{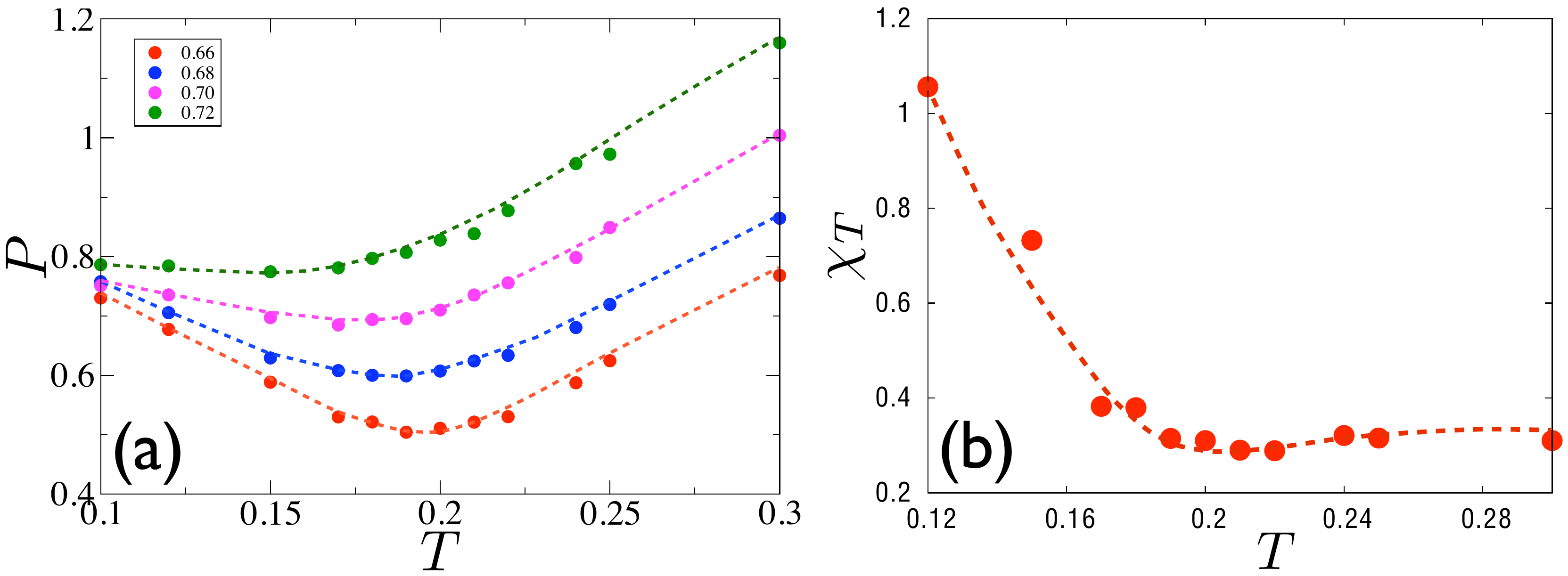}
\end{center}
\caption{(color-online)~(a) $P(T)$ isochores at $\rho = 0.66$(red), $0.68$(blue),
 $0.70$(pink) and $0.72$(green). Dashed lines are drawn as a guides to the eye. (b) The isothermal
 compressibility $\chi_T$ at $\rho = 0.67$. Note that the minimum in the pressure as well as in the
 compressibility is very similar to many common network forming liquids.}
\label{Fig.6}
\end{figure}

Networked liquids which coexist with solids of lower density show anomalous behaviour in many thermodynamic quantities. To investigate this we plot pressure-temperature curves along different {\em iso-chores} in Fig.\ref{Fig.6}(a) and each of these curves show a minimum. A minimum in the pressure iso-chore is equivalent to a maximum in density along an isobar as can be easily deduced as follows\cite{panomaly}. Noting that
\begin{equation}
\Big(\frac{\partial P}{\partial T}\Big)_{V} = \frac{\alpha_{P}}{\chi_{T}} \nonumber \\
\end{equation}
where $\chi_{T}$ is the isothermal compressibility, $\alpha_{P}$ is the volume-expansion coefficient, it is easy to see that an extremum in the isochore is related to vanishing $\alpha_{P}$. To determine the nature of the extremum we consider the second derivative, 
\begin{eqnarray}
\Big(\frac{\partial^{2} P}{\partial T^{2}}\Big)_{V} & = & \frac{\partial}{\partial T}\Big(\frac{\alpha_{P}}{\chi_{T}}\Big) \nonumber \\
 & = & -\frac{1}{\chi_{T}^{2}}\Big(\alpha_{P}\frac{\partial \chi_{T}}{\partial T} - \chi_{T}\frac{\partial \alpha_{P}}{\partial T}\Big) \nonumber \\
 & = & \frac{1}{\chi_{T}}\Big(\frac{\partial \alpha_{P}}{\partial T}\Big)
\end{eqnarray}
Thus, a density maximum (positive $\partial \alpha_P/\partial T$) leads to a pressure minimum. Further, $\chi_T$ itself may show an extremum at the point of minimum pressure, a fact which is borne out in our system and shown in Fig.\ref{Fig.6} (b).

\subsection{Density correlation functions}
Thermodynamic anomalies arise in a liquid as a result of a metastable liquid-liquid phase separation between a low density liquid with strong directional correlations and a more isotropic high density variant at temperatures where the solid phase is stable\cite{-vepressure1,-vepressure2}. One therefore expects that as the temperature is reduced, the homogeneous liquid would develop short ranged correlations which are strongly orientation dependent. In order to investigate this in our system, we have computed density-density correlation functions both from our simulations and from a liquid-state integral equation theory.
\begin{figure}[ht]
\begin{center}
\includegraphics[width=8.0cm]{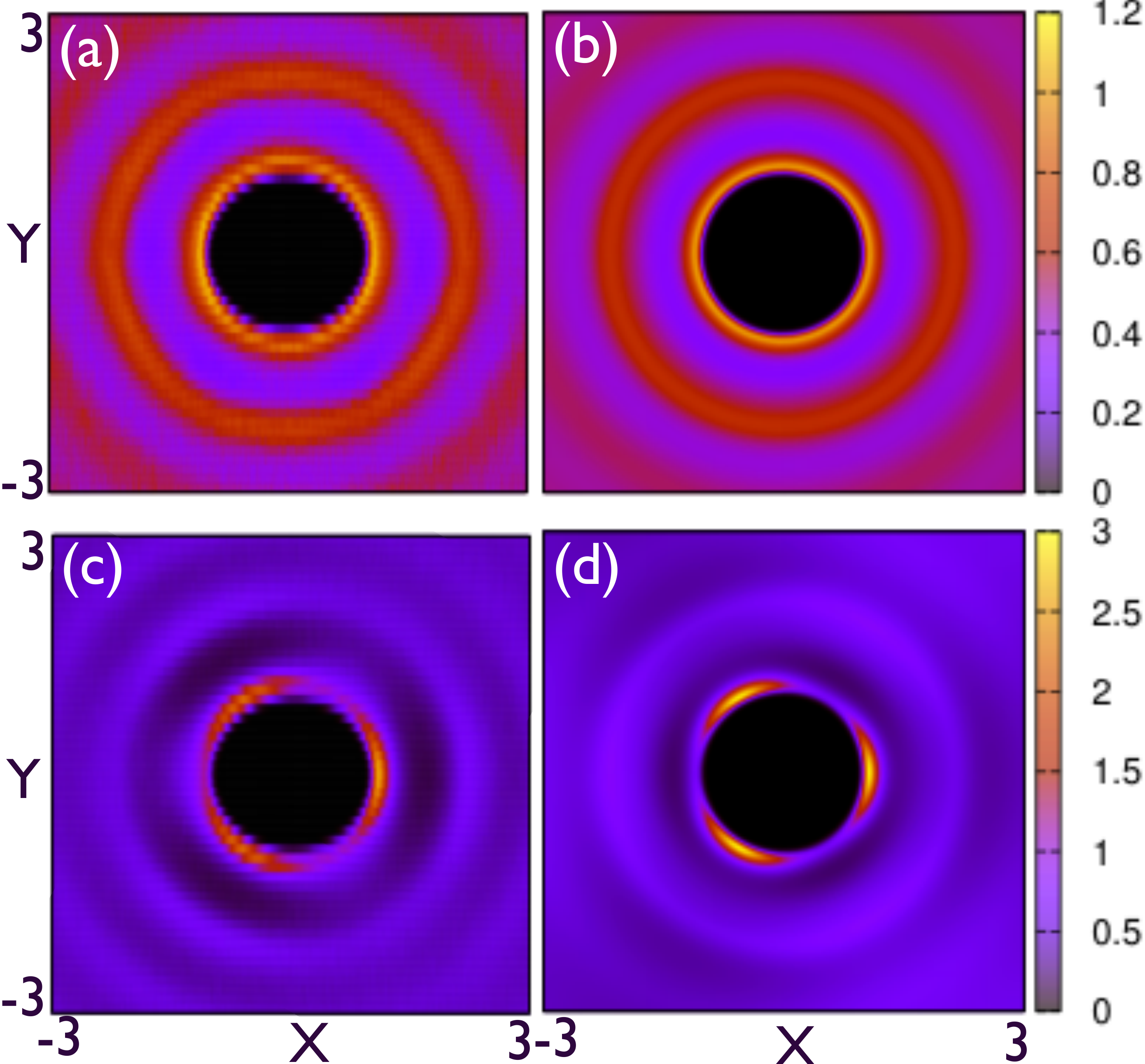}
\end{center}
\caption{(color-online)~Radial distribution functions $g_{1,1}({\bf r})$ at $T = 0.5$ obtained from (a) simulations and (b) theory and $g_{1,-1}({\bf r})$ at the same temperature obtained from (c) simulations and (d) theory.}
\label{Fig.7}
\end{figure}
In two dimensions, the radial-distribution function $g({\bf r})$ is defined as the probability of finding a pair of particles within ${\bf r}$ and ${\bf r} + {\bf dr}$ of each other given that one of particles is at the origin. In our case, we need distribution functions for both like, $g_{1,1}({\bf r})$, and unlike, $g_{1,-1}({\bf r})$, values of the internal coordinate. Similar to the pair interaction, a permutation of the indices $S_i$ in $g_{S_{i}S_{j}}({\bf r})$ results in a rotation in space by $\pi$, so that the functions $g_{1,-1}({\bf r})$and $g_{-1,1}({\bf r})$ are related by this transformation. We obtain these distribution functions at various temperatures and densities in the region where the liquid phase is stable by averaging over uncorrelated configurations. While $g_{1,1}({\bf r})$ is relatively insensitive to temperature, the nature of $g_{1,-1}({\bf r})$ depends strongly on $T$. We have illustrated this in Fig.\ref{Fig.7}. At low temperatures, when the system is dominated by it's potential energy which has a $3$-fold symmetry, $g_{1,-1}({\bf r})$ becomes large in magnitude compared to $g_{1,1}({\bf r})$, long ranged and shows strong directionality pointing to the formation of a prominent short ranged network in the liquid phase. This network is three fold coordinated similar to the HS phase and disappears when the temperature is increased.

In order to further understand our results, we compare the pair distribution functions obtained from our simulations with the results of an approximate integral equation theory which we describe below. Unlike a molecular fluid, in our case we need to set up equations for just two functions corresponding to like and unlike values of $S_{i}$ as in a binary mixture. The correlation functions are however direction dependent. In this case, we devise a perturbative scheme, where the direction dependent part is treated as a ``small'' perturbation over a set of isotropic functions. This allows us to quickly compute distribution functions which are in agreement with those obtained from simulations at high temperatures, but begins to deviate as the temperature is lowered. Nevertheless, our simple scheme is sufficient to show the emergence of the short-ranged directional order indicating network formation.

To begin, consider the Ornstein-Zernike equation for a binary mixture in Fourier space, viz.
\begin{equation}
\tilde{h}_{\alpha,\beta} ({\bf k}) = \tilde{c}_{\alpha,\beta}({\bf k}) + X_{\gamma} \tilde{c}_{\alpha,\gamma}({\bf k}) \tilde{h}_{\gamma,\beta}({\bf k})
\label{OZ}
\end{equation}
where the indices $\alpha, \beta = 1,-1$ denote two species of particles, $h_{\alpha,\beta} = g_{\alpha,\beta} - 1$ are the pair correlation functions, $X_k$ are the concentrations of species $k$ ($1/2$ in our case), and $c_{\alpha,\beta}$ are the direct correlation functions. Note that we have suppressed the spatial coordinate for simplicity.

In order to be useful, Eq.\ref{OZ} needs to be supplemented (or 'closed') with another equation involving the unknown functions $h_{\alpha,\beta}$ and $c_{\alpha,\beta}$. For short-ranged potentials an approximate closure relation which is known to work well is the Percus~-Yevick (PY) closure given by,
\begin{equation}
c_{\alpha,\beta}({\bf r}) = \exp[\frac{-U_{\alpha,\beta}({\bf r})}{k_{B}T}](1 + y_{\alpha,\beta}({\bf r})) - y_{\alpha,\beta}({\bf r}) - 1
\label{PY}
\end{equation}
with $y_{\alpha,\beta} = h_{\alpha,\beta} - c_{\alpha,\beta} $ the {\em indirect} correlation function and $U_{\alpha,\beta}({\bf r})$ the pair potential. We shall show that within our perturbative scheme, all the correlation functions may also be decomposed in the same way as $U_{\alpha,\beta}$ into isotropic and $\theta$ dependent parts as given in section 2A, namely,
\begin{eqnarray}
c_{1,1} = c_{-1,-1} & = & c_{0} (r) \nonumber \\
c_{1,-1} & = & c_{0}'(r) + \delta c(r,\theta) \nonumber \\
c_{-1,1} & = & c_{0}'(r) - \delta c(r,\theta)
\end{eqnarray}
and similarly for $y$.
\begin{figure}[ht]
\begin{center}
\includegraphics[width=9.0cm]{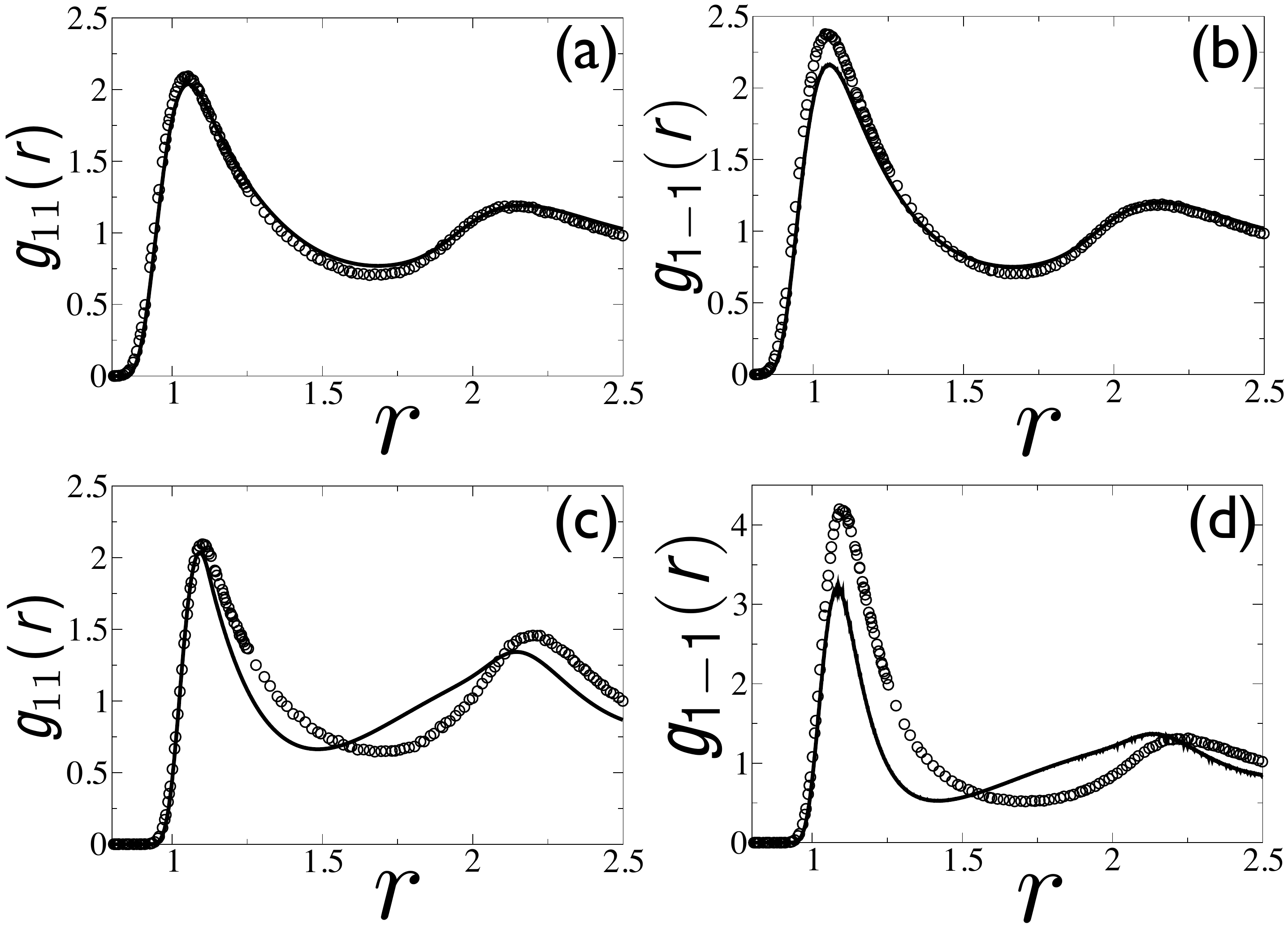}
\end{center}
\caption{~ Data obtained from  theory(solid curves) and simulations(open circles) are plotted together for $g_{1,1}(r)$ for (a) $T = 3.0$, (c) $T = 0.5$ and $g_{1,-1}(r)$ for (b) $T = 3.0$, (d) $T = 0.5$.}
\label{Fig.8}
\end{figure}
It is straight forward to show, after some algebra, that Eqn.\ref{OZ} reduces to,
\begin{eqnarray}
\tilde{y}_{0} & = &\frac{\tilde{c}_{0}\Delta +\frac{1}{2}(\tilde{c}_0^2 + \tilde{c}_0'^2)}{1 - \tilde{c}_{0} - \Delta} \nonumber \\
\tilde{y}_{0}^{'} & = & \frac{(\tilde{c}_{0}+\Delta)\tilde{c}_{0}^{'}}{1 - \tilde{c}_{0} - \Delta} \nonumber \\
\delta \tilde{y} & = & \frac{(\tilde{c}_{0}+\Delta)\delta \tilde{c}}{1 - \tilde{c}_{0} - \Delta} \nonumber \\
\label{OZ2}
\end{eqnarray} 
where $\Delta = \frac{1}{4}(\tilde{c}_0'^2 - \tilde{c}_0^2)$ and we have kept only terms up-to leading order in $\delta \tilde{c}$. The PY closure, similarly translates to,
\begin{eqnarray}
c_{0} & = & \exp \Big(\frac{-V_{0}}{k_B T}\Big)(1 + y_{0}) - (1 + y_{0}) \nonumber \\
c_{0}^{'} & = & (1+y_{0}^{'})[e^{-\beta(V_{0}')}\cosh(\beta \delta V) - 1] \nonumber \\
& & - (\delta y)e^{-\beta(V_{0}')}\sinh(\beta \delta V) \nonumber \\
\delta c & = & (\delta y)[e^{-\beta(V_{0}')}\cosh(\beta \delta V) - 1] \nonumber \\
& & - (1+y_{0}^{'})e^{-\beta(V_{0}')}\sinh(\beta \delta V) \nonumber \\
\label{PY2}
\end{eqnarray}
A further, ad-hoc, approximation simplifies the problem considerably; the aposteriori justification being given by its ability to reproduce some of the more essential features of the correlation functions as compared with the output of our simulations. Accordingly, we take $\delta V \to 0$ in the expression for  $c_0'$ in Eq.\ref{PY2}. In this case the Eqns. \ref{OZ} and \ref{PY} factorize into isotropic and $\theta$ dependent parts. We solve the resulting integral equations for the isotropic functions self-consistently and then calculate the $\theta$ dependendent pair correlation functions by iterating Eq.\ref{PY2} once.  
Our results are compared in Figs.\ref{Fig.7} and \ref{Fig.8}. In spite of the approximations made, we find fair agreement between simulations and our integral equation theory especially at high temperatures. In Fig.\ref{Fig.7} (a) and (b) we show plots of the radial distribution function $g_{1 1}({\bf r})$ obtained from simulations and solution of the integral equations respectively. None of them show any strong directional dependence. On the other hand a similar plot of $g_{1 -1}$ shows a very strong three fold directional dependence pointing out the emergence of a local three fold coordinated, short-ranged network structure. In Fig.\ref{Fig.8} we plot the $\theta$-averaged radial distribution functions $g_{1,1}(r)$ and $g_{1,-1}(r)$ from theory and simulations both at high and low temperature. 

While at high temperature Fig.\ref{Fig.8}(a)and(b) the pair distribution function does not show any preference for either like or unlike species, one observes in both the simulation data as well as the solutions to the integral equations, a strong preference for unlike oriented neighbors emerging as the temperature is reduced Fig.\ref{Fig.8}(c)and(d), due to the development of a local network. Since the potential energy and hence the $\theta$-dependent part has a significant effect on the system at low temperatures, the radial distribution functions obtained from theory and simulation however do not match quantitatively in the low-temperature limit. More specifically, at temperatures where our integral equation are valid, the local network is not strong enough to give rise to thermodynamic anomalies which occur for cooler liquids. Better approximations by keeping orientation dependent terms to all orders are needed to yield accurate correlation functions in the desired range at the cost of substantially increasing the computational complexity.

\section{Conclusions and future directions}
We have introduced a model-system in two-dimensions to study whether network formation in a system leads to polymorphism and thermodynamic anomalies even if molecular rotational degrees of freedom are not explicitly taken into account. While the connection between directional bonding and the existence of multiple states with differences in density which causes density anomalies has been established quite generally\cite{lattice-model}, our model shows that these properties are robust against rather drastic simplifications of the nature of the rotational states. We study the equilibrium properties of this system in detail and show that, though simple, the model shows many features of real liquids like water.

Our studies should be of direct relevance to the biologically important case of confined water\cite{bagchi} where rotations are strongly coupled to translational degrees of freedom not unlike the case studied here. Indeed, we expect strong network formation in such cases and the liquid should show prominent thermodynamic anomalies. Calculations in this direction are in progress and will be published elsewhere. In future, we aim to use our model in more complicated situations to discuss issues such as shear flow and coupling to external fields to further study the properties of confined network formers.

\acknowledgements
Support from the Indo-Eu project MONAMI is acknowledged. C.M. acknowledges support from a CSIR Senior Research Fellowship. Illuminating discussions with S. Sastry are gratefully acknowledged.


\begin{thebibliography}{99}
\bibitem{torque} Note: Since the forces on the particles are angle dependent, the total torque on the system is non-zero. We set this torque to be zero, however, since it does not affect the equilibrium properties.
\end{thebibliography}
\end{document}